\RequirePackage{etoolbox}

\newbool{JINST}
\boolfalse{JINST}

\ifbool{JINST}{
    \documentclass[11pt,a4paper]{article}
    \usepackage{jinstpub}
}{
    \documentclass[12pt,a4paper]{article}

    \setlength{\hoffset}{-2cm}
    \setlength{\voffset}{-2cm}
    \topmargin=0.5cm
    \oddsidemargin=2.5cm
    \evensidemargin=-10mm
    \textwidth=16cm
    \textheight=230mm
    \headsep=20mm
    \columnsep=5mm

    \raggedbottom
    \sloppy

    \usepackage{cite}
    \def\note#1{{\color{red} Note: #1 \color{black}}}
}

\usepackage{amsmath}
\usepackage{amssymb}
\usepackage{amsfonts}
\usepackage{upgreek}
\usepackage[T1]{fontenc}
\usepackage{fixltx2e}
\usepackage{lmodern}
\usepackage{textcomp}
\usepackage{gensymb}

\usepackage{epsfig}
\usepackage{graphicx}
\graphicspath{{./figs/}} 
\usepackage{xspace}
\usepackage{enumerate}
\usepackage{multirow}

\usepackage{parskip}

\usepackage{sidecap}
\sidecaptionvpos{figure}{c}

\usepackage{hyperref}
\usepackage{subfig}
\usepackage{color}
\definecolor{LightBlue}{RGB}{70,130,180}
\usepackage{float}
\usepackage{appendix}
\usepackage{listings}
\usepackage{booktabs}
\usepackage{dcolumn}
\newcolumntype{.}{D{.}{.}{-1}}
\usepackage{ifthen} 
\newboolean{uprightparticles}
\setboolean{uprightparticles}{true}
\newboolean{articletitles}
\setboolean{articletitles}{true} 

\def\gtap{\mathrel{\rlap{\raise 0.511ex \hbox{$>$}}{\lower 0.511ex\hbox{$\sim$}}}} 

\usepackage{mciteplus}

\def\dut {\mbox{DUT}\space}

\def\ux85 {\mbox{UX85}\xspace}

\def\tpxt   {\mbox{Timepix3}\xspace}

\ifthenelse{\boolean{uprightparticles}}%
{

 \def\PDelta      {\ensuremath{\Delta}\xspace}                 
 \def\PXi      {\ensuremath{\Xi}\xspace}                 
 \def\PLambda      {\ensuremath{\Lambda}\xspace}                 
 \def\PSigma      {\ensuremath{\Sigma}\xspace}                 
 \def\POmega      {\ensuremath{\Omega}\xspace}                 
 \def\PUpsilon      {\ensuremath{\Upsilon}\xspace}                 
 
 \def\PB      {\ensuremath{\mathrm{B}}\xspace}                 
                  
 \def\PD      {\ensuremath{\mathrm{D}}\xspace}

 \def\PK      {\ensuremath{\mathrm{K}}\xspace}

 \def\Pi      {\ensuremath{\mathrm{i}}\xspace}

}
{

 \mathchardef\PDelta="7101
 \mathchardef\PXi="7104
 \mathchardef\PLambda="7103
 \mathchardef\PSigma="7106
 \mathchardef\POmega="710A
 \mathchardef\PUpsilon="7107
                  
 \def\PB      {\ensuremath{B}\xspace}                 
                  
 \def\PD      {\ensuremath{D}\xspace}

 \def\PK      {\ensuremath{K}\xspace}

 \def\Pi      {\ensuremath{i}\xspace}

}

\def\kaon  {\ensuremath{\PK}\xspace}
  \def\Kbar  {\kern 0.2em\overline{\kern -0.2em \PK}{}\xspace}

\def\Kz    {\ensuremath{\kaon^0}\xspace}
\def\Kzb   {\ensuremath{\Kbar^0}\xspace}
\def\KzKzb {\ensuremath{\Kz \kern -0.16em \Kzb}\xspace}
\def\Kp    {\ensuremath{\kaon^+}\xspace}
\def\Km    {\ensuremath{\kaon^-}\xspace}

\def\KpKm  {\ensuremath{\Kp \kern -0.16em \Km}\xspace}

  \def\Dbar    {\kern 0.2em\overline{\kern -0.2em \PD}{}\xspace}
\def\D       {\ensuremath{\PD}\xspace}

\def\Dz      {\ensuremath{\D^0}\xspace}
\def\Dzb     {\ensuremath{\Dbar^0}\xspace}
\def\DzDzb   {\ensuremath{\Dz {\kern -0.16em \Dzb}}\xspace}
\def\Dp      {\ensuremath{\D^+}\xspace}
\def\Dm      {\ensuremath{\D^-}\xspace}

\def\DpDm    {\ensuremath{\Dp {\kern -0.16em \Dm}}\xspace}

\def\Bbar    {\ensuremath{\kern 0.18em\overline{\kern -0.18em \PB}{}}\xspace}

  \def\Y#1S{\ensuremath{\PUpsilon{(#1S)}}\xspace}

\def\Lbar {\ensuremath{\kern 0.1em\overline{\kern -0.1em\PLambda}}\xspace}

\def\AT#1     {\ensuremath{A_{\mathrm{T}}^{#1}}\xspace}           

\def\C#1      {\ensuremath{\mathcal{C}_{#1}}\xspace}                       
\def\Cp#1     {\ensuremath{\mathcal{C}_{#1}^{'}}\xspace}                    
\def\Ceff#1   {\ensuremath{\mathcal{C}_{#1}^{\mathrm{(eff)}}}\xspace}        
\def\Cpeff#1  {\ensuremath{\mathcal{C}_{#1}^{'\mathrm{(eff)}}}\xspace}       
\def\Ope#1    {\ensuremath{\mathcal{O}_{#1}}\xspace}                       
\def\Opep#1   {\ensuremath{\mathcal{O}_{#1}^{'}}\xspace}                    

\newcommand{\unit}[1]{\ensuremath{\rm\,#1}\xspace}          

\newcommand{\tev}{\ensuremath{\mathrm{\,Te\kern -0.1em V}}\xspace}
\newcommand{\gev}{\ensuremath{\mathrm{\,Ge\kern -0.1em V}}\xspace}
\newcommand{\mev}{\ensuremath{\mathrm{\,Me\kern -0.1em V}}\xspace}
\newcommand{\kev}{\ensuremath{\mathrm{\,ke\kern -0.1em V}}\xspace}
\newcommand{\ev}{\ensuremath{\mathrm{\,e\kern -0.1em V}}\xspace}
\newcommand{\gevc}{\ensuremath{{\mathrm{\,Ge\kern -0.1em V\!/}c}}\xspace}
\newcommand{\mevc}{\ensuremath{{\mathrm{\,Me\kern -0.1em V\!/}c}}\xspace}
\newcommand{\gevcc}{\ensuremath{{\mathrm{\,Ge\kern -0.1em V\!/}c^2}}\xspace}
\newcommand{\gevgevcccc}{\ensuremath{{\mathrm{\,Ge\kern -0.1em V^2\!/}c^4}}\xspace}
\newcommand{\mevcc}{\ensuremath{{\mathrm{\,Me\kern -0.1em V\!/}c^2}}\xspace}

\def\mum  {\ensuremath{\,\upmu\rm m}\xspace}

\def\mus  {\ensuremath{\,\upmu{\rm s}}\xspace}
\def\ns   {\ensuremath{{\rm \,ns}}\xspace}
\def\ps   {\ensuremath{{\rm \,ps}}\xspace}

\def\gsim{{~\raise.15em\hbox{$>$}\kern-.85em
          \lower.35em\hbox{$\sim$}~}\xspace}
\def\lsim{{~\raise.15em\hbox{$<$}\kern-.85em
          \lower.35em\hbox{$\sim$}~}\xspace}

\def\urad{\ensuremath{\,\upmu\rm rad}\xspace}

\def\gaudi      {\mbox{\textsc{Gaudi}}\xspace}
\def\kepler     {\mbox{\textsc{Kepler}}\xspace}

\def\tell1  {TELL1\xspace}
\def\ukl1   {UKL1\xspace}

\begin{document}


\ifbool{JINST}{

    \title{LHCb VELO Timepix3 Telescope}

    \author[a]{Kazu~Akiba,}
\author[a,1]{Martin~van~Beuzekom%
\note{Corresponding author.},}
\author[a]{Henk~Boterenbrood,}
\author[b]{Emma~Buchanan,}
\author[c]{Jan~Buytaert,}
\author[c]{Wiktor~Byczynski,}
\author[d]{Xabier~Cid~Vidal,}
\author[c]{Paula~Collins,}
\author[a]{Elena~Dall'Occo,}
\author[d]{Alvaro~Dosil~Su\'{a}rez,}
\author[c]{Raphael~Dumps,}
\author[e]{Tim~Evans,}
\author[f]{Vinicius~Franco~Lima,}
\author[d]{Abraham~Gallas~Torreira,}
\author[d]{Juli\'{a}n~Garc\'{i}a~Pardi\~{n}as,}
\author[a]{Bas~van~der~Heijden,}
\author[g]{Christoph~Hombach,}
\author[e]{Malcolm~John,}
\author[c]{Szymon~Kulis,}
\author[c]{Xavi~Llopart~Cudie,}
\author[h]{Franciole~Marinho,}
\author[b]{Eugenia~Price,}
\author[b]{Sophie~Richards,}
\author[g]{Pablo~Rodriguez~Perez,}
\author[b]{Daniel~Saunders,}
\author[c]{\r{A}smund~Schiager~Folkestad,}
\author[c]{Heinrich~Schindler,}
\author[a]{Frans~Schreuder,}
\author[a]{Hella~Snoek,}
\author[a]{Panagiotis~Tsopelas,}
\author[b]{Jaap~Velthuis,}
\author[d]{Maria~Vieites~Diaz,}
\author[g]{Mark~J.~R.~Williams}

\affiliation[a]{Nikhef, Science Park 105, 1098 XG Amsterdam, the Netherlands}
\affiliation[b]{University of Bristol, Beacon House, Queens Road,  BS8 1QU, Bristol,United Kingdom}
\affiliation[c]{CERN, 1211 Geneve, Switzerland}
\affiliation[d]{IGFAE, University of Santiago de Compostela, Rúa de Xoaquín Díaz de Rábago, 15782 Santiago de Compostela, Spain}
\affiliation[e]{University of Oxford, Particle Physics Department Denys Wilkinson Bldg., Keble Road, Oxford OX1 3RH, United Kingdom }
\affiliation[f]{Liverpool University, Oliver Lodge Laboratory, Oxford Street, Liverpool L69 7ZE, United Kingdom }
\affiliation[g]{School of Physics and Astronomy, University of Manchester, Oxford Rd, Manchester M13 9PL, United Kingdom }
\affiliation[h]{Federal University of S\~ao Carlos,  Rodovia Anhanguera, km 174, 13604-900 Araras,  Brazil}

\emailAdd{m.van.beuzekom@nikhef.nl}

    \abstract{
    The LHCb VELO Timepix3 telescope is a silicon pixel tracking system
constructed initially to evaluate the performance of LHCb VELO Upgrade prototypes.
The telesope consists of eight hybrid pixel silicon sensor planes equipped with the Timepix3 ASIC. The planes
provide excellent charge measurement, timestamping and spatial resolution and the system can
function at high track rates.  This paper describes the construction of the telescope and its data acquisition
system and offline reconstruction software.
A timing resolution of 350~ps was obtained for reconstructed tracks.
A pointing resolution of better than 2~\mum was determined  for the 180~GeV/c 
mixed hadron beam at the CERN SPS. The telescope has been shown to operate at
a rate of 5 million particles~\unit{s^{-1}\cdot cm^{-2}}
without a loss in efficiency.

    }

}{

    \begin{titlepage}



    {\bf\huge
    \begin{center}
      LHCb VELO Timepix3 Telescope
    \end{center}
    }

    {\bf\boldmath\huge
    \begin{center}
    \end{center}
    }


    \begin{center}
    \begin{center}
\small

 Kazu~Akiba$^{a}$, 
 Martin~van~Beuzekom$^{a,1}$, 
 Henk~Boterenbrood$^{a}$, 
 Emma~Buchanan$^{b}$,
 Jan~Buytaert$^{c}$,
 Wiktor~Byczynski$^{c}$,
 Xabier~Cid~Vidal$^{d}$,
 Paula~Collins$^{c}$,
 Elena~Dall'Occo$^{a}$, 
 Alvaro~Dosil~Su\'{a}rez$^{d}$,
 Raphael~Dumps$^{c}$,
 Tim~Evans$^{e}$,
 Vinicius~Franco~Lima$^{f}$,
 Abraham~Gallas~Torreira$^{d}$,
 Juli\'{a}n~Garc\'{i}a~Pardi\~{n}as$^{d}$,
 Bas~van~der~Heijden$^{a}$, 
 Christoph~Hombach$^{g}$,
 Malcolm~John$^{e}$,
 Szymon~Kulis$^{c}$,
 Xavi~Llopart~Cudie$^{c}$,
 Franciole~Marinho$^{h}$,
 Eugenia~Price$^{b}$,
 Sophie~Richards$^{b}$,
 Pablo~Rodriguez~Perez$^{g}$,
 Daniel~Saunders$^{b}$,
 \r{A}smund~Schiager~Folkestad$^{c}$,
 Heinrich~Schindler$^{c}$,
 Frans~Schreuder$^{a}$, 
 Hella~Snoek$^{a}$, 
 Panagiotis~Tsopelas$^{a}$, 
 Jaap~Velthuis$^{b}$,
 Maria~Vieites~Diaz$^{d}$,
 Mark~R.~J.~Williams$^{g}$

\smallskip{\it\footnotesize $ ^{a}$Nikhef \\ }
\smallskip{\it\footnotesize $ ^{b}$Bristol University\\ }
\smallskip{\it\footnotesize $ ^{c}$CERN\\ }
\smallskip{\it\footnotesize $ ^{d}$University of Santiago de Compostela \\ } 
\smallskip{\it\footnotesize $ ^{e}$Oxford University \\ }
\smallskip{\it\footnotesize $ ^{f}$Liverpool University\\ }
\smallskip{\it\footnotesize $ ^{g}$University of Manchester\\ }
\smallskip{\it\footnotesize $ ^{h}$Universidade Federal de S\~ao Carlos\\ }
\smallskip{$ ^{1}$ Corresponding author: m.van.beuzekom@nikhef.nl }

\end{center}

    \end{center}

    \vspace*{2.0cm}

    \begin{abstract}
    \noindent
    The LHCb VELO Timepix3 telescope is a silicon pixel tracking system
constructed initially to evaluate the performance of LHCb VELO Upgrade prototypes.
The telesope consists of eight hybrid pixel silicon sensor planes equipped with the Timepix3 ASIC. The planes
provide excellent charge measurement, timestamping and spatial resolution and the system can
function at high track rates.  This paper describes the construction of the telescope and its data acquisition
system and offline reconstruction software.
A timing resolution of 350~ps was obtained for reconstructed tracks.
A pointing resolution of better than 2~\mum was determined  for the 180~GeV/c 
mixed hadron beam at the CERN SPS. The telescope has been shown to operate at
a rate of 5 million particles~\unit{s^{-1}\cdot cm^{-2}}
without a loss in efficiency.

    \end{abstract}

    
    \end{titlepage}
    
    \pagestyle{empty}  
    

    \setcounter{page}{2}
    
}

\ifbool{JINST}{
    \maketitle
}{ 
    \tableofcontents
    \pagestyle{plain}
    \setcounter{page}{1}
    \pagenumbering{arabic}
    \clearpage
}

\section{Introduction}
Hybrid pixel detectors from the Medipix~\cite{medipixcolab} ASIC family have been previously employed for reconstruction of trajectories of high energy charged particles~\cite{tpxtelepaper1, tpxtelepaper2}. The combination of powerful integrated circuits and the high signal to noise ratio of the sensor assemblies makes
this technology ideal for precise track reconstruction. 
The Timepix3 ASIC~\cite{tpx3asic} offers very precise and simultaneous measurements of the particle's Time-of-Arrival (ToA) and Time-over-Threshold (ToT), which is proportional to the deposited charge of the corresponding signal.
These two measurements provide an excellent input for the pattern recognition algorithms, making the subsequent 
track reconstruction fast and precise.

This paper describes a high performance particle telescope which has been constructed based on the Timepix3
ASIC, in the view of the sensor characterisation program for the upgrade of the LHCb Vertex Locator (VELO)~\cite{veloupgrade}.
The telescope reconstructs timestamped tracks with a precise projected position in the centre of the
telescope.  A Device-Under-Test (DUT) can be mounted at this position and moved via remote-controlled 
rotation and translation stages.  The telescope has already integrated a variety of DUTs in the
telescope data acquisition system; either with an identical readout in the case of Timepix3 assemblies, or via
offline timestamp alignment.
To characterise DUT sensors in detail, charged particle beams at the PS and SPS~\cite{PS,SPS}
were used. These facilities provide protons with momenta of 24 GeV/c (PS) and 
mixed hadrons (p, $\pi$, K) of about 120 GeV/c (SPS). Most of the results presented 
are obtained with SPS hadrons that are provided in spills of about 2 million particles
during 4.5 seconds.

The telescope mechanical design is described in section~\ref{mech}, followed by
the readout chain and slow controls in sections~\ref{daq} and \ref{slow}, and the data analysis software packages in 
section~\ref{offline}. 
Finally, the relevant telescope performance figures required to assess the quality of DUTs are described in section~\ref{performance}.

\section{Hardware Description}\label{mech}
\begin{figure}
  \centering
  \includegraphics[width=0.9\textwidth]{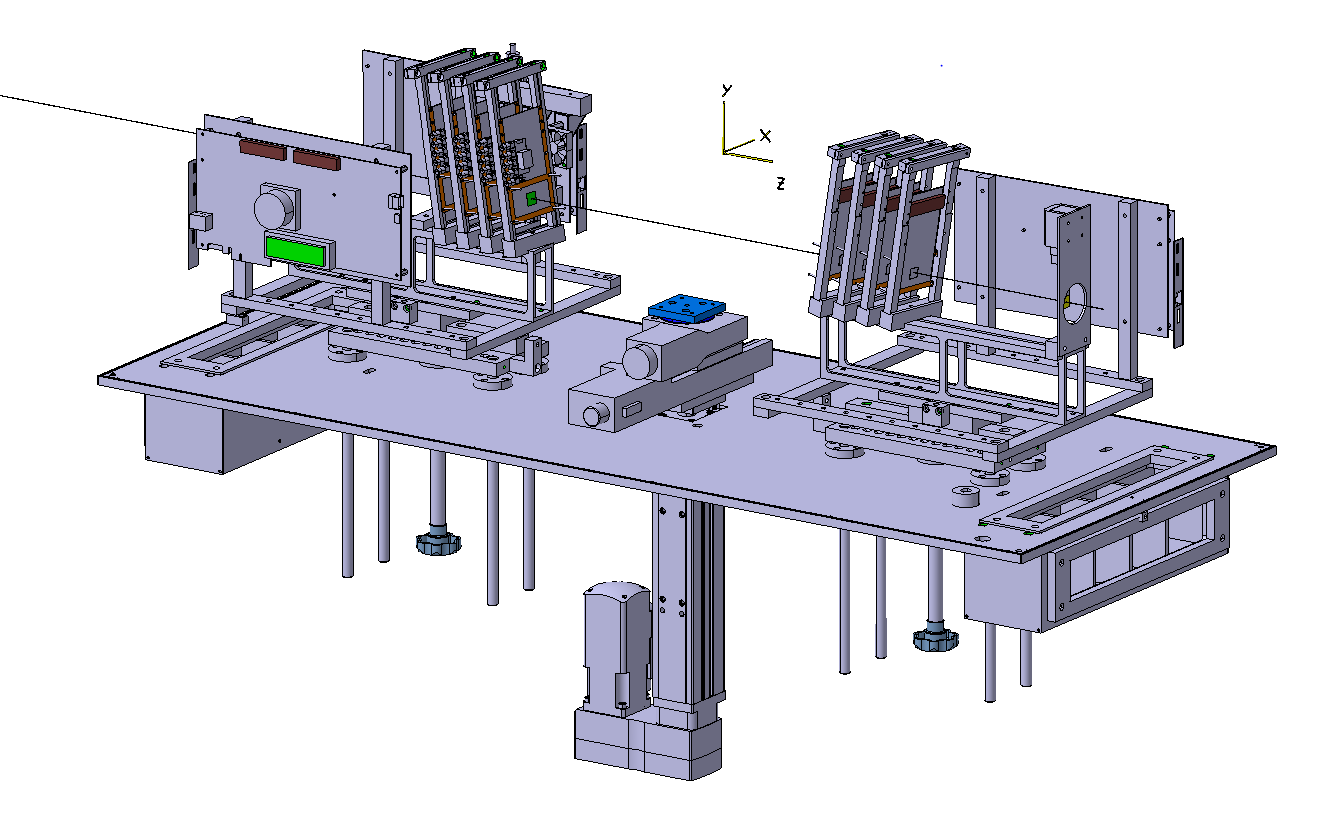}
  \caption{Mechanical design of the Timepix3 telescope, with the coordinate system
 displayed at the top. The telescope stations are mounted on two retractable arms 
around a central stage.  The central stage is  reserved for studies on DUTs; 
it provides translations in $x$ and $y$  as well as rotations about the $y$ axis.   }
  \label{Fig:DrawingTelescope}
\end{figure}

The telescope consists of two arms of four detector planes each, as 
illustrated in figure~\ref{Fig:DrawingTelescope}. 
A global right-handed coordinate frame is defined  
with the $z$ axis in the direction of the beam and the 
$y$ axis pointing upwards. This convention is adopted throughout this paper. 
The detector modules are 300\,\mum thick $p$-on-$n$ pixelised silicon sensors bump-bonded to 
$\approx 700$\,\mum thick Timepix3 ASICs.
Hence, each plane adds approximately 1~mm of silicon 
($\approx 1\%~X_{0}$, where $X_{0}$ is the radiation length) to the material budget of the telescope. 
The printed circuit boards that hold the sensor assemblies have a square cut-out beneath the 
active area of the chip and thus do not contribute to the material budget, 
except for an outer ring of about 0.5~mm.
To improve the spatial resolution, charge sharing between pixels 
is maximised by 
placing the planes under an angle 
about the  $x$ and $y$ with respect to the $z$~axis.
For 300~$\mum$ thick sensors and 55~\mum pixel pitch the optimal angle corresponds to 
approximately  $9^\circ$.
Given the pixel pitch and a matrix size of 256$\times$256, 
the active area of the telescope is about 1.4$\times$1.4~cm$^2$. 
The positions of the telescope planes along the $z$ axis are adjustable;
the telescope is operated typically with a distance of 31~mm between the planes within one arm and about 200
mm between the two arms.
Cooling of the telescope planes is achieved with pre-cooled air circulating inside  the
telescope protective enclosure, which is also light-tight.

A \dut can be installed on a remote-controlled motion 
stage between the two telescope arms.
The two telescope arms are mounted on rails such that the distance 
between the arms can be adjusted depending on the size of the DUT.
The stage allows translations in $x$ and $y$ and rotations about the $y$ 
axis. 
A second stage is available downstream of the telescope which is
used by DUTs larger than the available space or less demanding in terms of  pointing resolution. 
The telescope also includes two scintillators, 
one upstream and one downstream of the telescope arms, which can be 
used to trigger the data acquisition systems of external 
DUTs\footnote{External devices are detectors that are not based on the Timepix3 ASIC.}.

Finally, the base of the telescope is mounted on a remote controlled  
motion stage, which allows the entire telescope to be moved in $x$ and $y$.
This allows the alignment of the telescope with respect to the beam.




\section{Data Acquisition (DAQ)}\label{daq}
The telescope silicon sensor signals are processed by Timepix3 ASICs that can be operated in
different readout  modes~\cite{tpx3asic}. The acquisition mode used for the telescope features
simultaneous measurement of the Time-of-Arrival (ToA) and Time-over-Threshold (ToT) per pixel.  A
pixel hit is timestamped with a Time-to-Digital converter bin size of
1.56~ns, and a timestamp range of 409.6~\mus.
The ToT measurement provides information about the amount of charge collected by the pixel.  It has
a range of 10 bits, with a configurable resolution that can be set via the discharge current of the
preamplifier.  The typical operational conditions for the telescope give a correspondence of each
ToT count to a charge of about 200~e$^-$.  The capacitance of the small pixels is low and hence the
noise contribution from the amplifier circuit is less than 100~e$^-$ RMS.  The shot noise contribution
from the sensor is negligible.  Running with a nominal threshold of 1000~e$^-$ for the telescope
planes ensures that there are virtually no noise hits, while allowing full detection efficiency for
multi pixel clusters.

For high rate tracking performance, the chip features a so-called data-driven readout mode in which
the 48 bit data packet of each pixel hit is sent off-chip immediately after ToT conversion.  The
chip can send out up to 80 Mhits/s over 8 serial links each running at 640 Mbit/s.  To acquire the
data from the ASICs a readout system called Speedy PIxel Detector Readout (SPIDR)~\cite{SPIDR,SPIDR2} is
employed.  The SPIDR system is specially tailored to the readout of the Timepix3 and Medipix3 chips
at their maximum rate, and can be operated with both a 1 Gigabit and 10 Gigabit ethernet (GbE)
interface.  The 10 GbE interface is $\approx 55\%$ occupied when reading out a single Timepix3 chip
at its maximum rate.  
The  SPIDR version  used in the telescope  is based on the Xilinx Virtex 7
VC707 evaluation board and can read out two telescope planes in parallel.  This is the default
configuration of the readout, but to operate the whole telescope at its maximum rate  each telescope
plane requires a single dedicated SPIDR board.  
 
Three data streams can be distinguished: slow control to and from the SPIDR via TCP/IP, 
the (full) data stream from SPIDR to computer via the User Datagram Protocol (UDP), and monitoring data from SPIDR to computer via UDP. 

The two main tasks of the Virtex 7 FPGA board are the slow and fast (clock synchronisation and timing) control of the Timepix3 chips, 
and the packing of the data coming from the chips into UDP datagrams. 
Upon reception of the pixel packets, the timestamp in the packet with a range of 409.6\mus 
is extended by 16 bits, thereby increasing the range to 28.8 seconds to ease the offline reconstruction.
Synchronisation of the different telescope planes is achieved via a central logic unit (TLU,
Telescope Logic Unit) that provides the clock to all SPIDR systems.  The TLU hardware is based on a
Xilinx SP601 evaluation board and the fan out of the signal is is done with boards originally
designed for a cosmic ray experiment~\cite{TLUfanout}.  The TLU also supplies a signal to
synchronise all time counters, and a shutter signal to synchronously control the start and stop of
the data flow.  Each SPIDR supplies a \emph{busy} signal to the TLU which is used to signal buffer
overflow either in the SPIDR or its DAQ computer.  If one of the systems overflows the shutter is
closed by the TLU and the entire data acquisition is paused until the buffers are emptied. 
The SPIDR boards and the TLU receive commands from the run control software via ethernet. The run
control software provides a graphical user interface for configuring the boards and ASICs and to
start and stop runs.  Moreover, the run control graphically displays the volume of data acquired per
telescope plane. 
In this way problems in the data acquisition can be monitored, which occasionally occur due to
radiation induced Single Event Upsets in the configuration registers of the Timepix3.

In order to simplify integration of other, non-Timepix3, detectors, the SPIDR boards feature a TDC
input with 260~ps bins.  The trigger signal of such a detector can then be timestamped in the clock
domain of the telescope and reconstructed tracks can be linked to these triggers in the offline
software. Since the telescope uses four SPIDR boards, up to four different detectors can be run
simultaneously with the telescope.

\section{Slow Control and Monitoring}\label{slow}

\begin{figure}[h]
  \centering
  \includegraphics[width=0.9\textwidth]{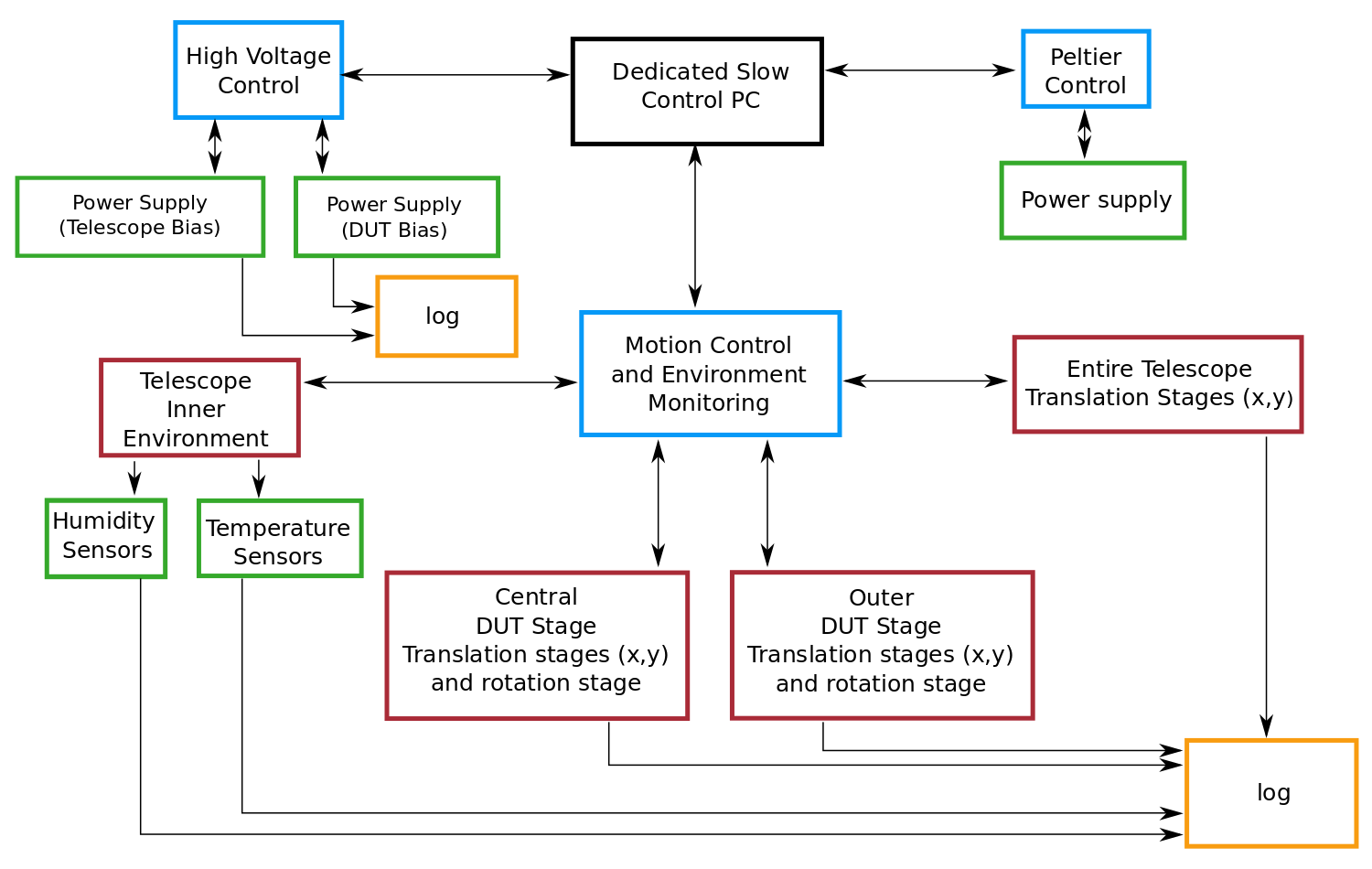}
  \caption{Block diagram of the slow control system. 
The telescope and DUT bias voltages and motion stages are operated remotely. 
The bias currents and temperatures of the telescope environment are monitored and logged. }
  \label{Fig:SlowControlBlock}
\end{figure}

A dedicated slow control system has been developed to enable remote-control of all motion stages and power supplies. 
The entire telescope can be moved in $x$ and $y$ to remotely align with the beam position, 
or be removed from the beam path completely if required. 
The two DUT stages, one mounted at the centre between the two telescope arms 
and the other mounted outside of the main telescope enclosure, 
include precise remote-controlled translation and rotation stages. 
The translation and rotation stages, supplied by PI~\cite{PI},  
are controlled by stepper motors and have a repeatability of 2~\mum for translation and 50~\urad for rotation.
The vertical stages are supplied by Festo~\cite{Festo}.
Two Keithley 2410 Source Meters~\cite{Keithley} are used to bias independently  the telescope planes and DUT.
The telescope planes are usually operated with a bias tension of 150 V, which is at least 100~V above the sensor full depletion voltage. 
An additional power supply is also controlled remotely  for a Peltier based DUT cooling system. 
Humidity and temperature sensors provide environmental monitoring of conditions within the telescope enclosure. 
The positions of all motion stages as well as the bias voltages, 
currents and environmental measurements are logged periodically to enable recovery of 
all conditions for analysis. Figure~\ref{Fig:SlowControlBlock} summarises all elements of the slow control system.

\section{Offline Reconstruction Software}\label{offline}

A software application, ``\kepler\footnote{in honour of the German astronomer Johannes Kepler (1571-1630).}'',
 based on the \gaudi event-processing framework~\cite{Clemencic2009}
has been developed to analyse the telescope data and provides track fitting,  alignment and 
charge calibration among other tasks. 
The application produces a ROOT file with histograms and ntuples for further user analyses.

The \gaudi framework was developed to process data taken at colliders, where there is a clear `event' 
structure corresponding to the bunch crossing. No such structure exists in the PS and SPS secondary beams, 
where there is a continuous time profile during the flat top of a spill, which 
typically lasts about 4.5 seconds for the CERN SPS.
In contrast to frame-based 
telescopes~\cite{tpxtelepaper2}, the \tpxt\ telescope data does not have an inherent event structure due to the 
data-driven readout described in section~\ref{daq}.
Because the typical dataset of a run is hundreds of MBytes it is impractical to process the entire run in a single pass.
The data stream is therefore divided into time slices of configurable length and each slice is processed individually.
The length of the time slice is chosen to be 400~\mus and with this setting the number of tracks that are split up into different events is negligible.

For each event, a sequence of algorithms is executed, 
starting by decoding the raw data files for each detector plane.
The pixel packets are not received in order with respect to their timestamps.
In order to ensure all hits belonging to the current event are found, it is 
necessary to read ahead in the file until a packet 
with a timestamp newer than the next-to-next event, 
i.e. over 800~\mus after the start of the current event, is found.  

The reconstruction of one spill ($\sim10^{6}$ tracks) takes approximately 
70 seconds on an Intel E3 2.6 GHz processor. 
The large-scale processing of runs on a distributed analysis platform such as 
DIRAC is enabled in GANGA \cite{Ganga}, a  process  management interface.


In the following sections the general structure of the telescope reconstruction software is described.
The ToT to charge calibration method is described in section~\ref{calib}, followed by the  time alignment in
section~\ref{timealignment}. The clustering and tracking algorithms are detailed in section~\ref{cluster}.
Finally the spatial alignment is described in section~\ref{alignment}.

\subsection{Charge calibration} \label{calib} 
The Timepix3 ToT measurement provides information about the charge deposited in the
sensor. In order to find the correspondence between ToT and charge, a calibration procedure
is performed using test pulses. 
The ToT values are converted to charge with a surrogate function~\cite{Jakubek2008}. The parameters of this 
function are obtained for each pixel by injecting known charges via a test-pulse circuit and measuring the 
corresponding ToT values. The calibration has been verified using monochromatic x-rays in the Brazilian
Synchrotron Light Laboratory. These two methods agree within 4\%~\cite{thesismateus}.

\begin{figure}[!htb] 
  \centering 
  \includegraphics[width=.60\textwidth]{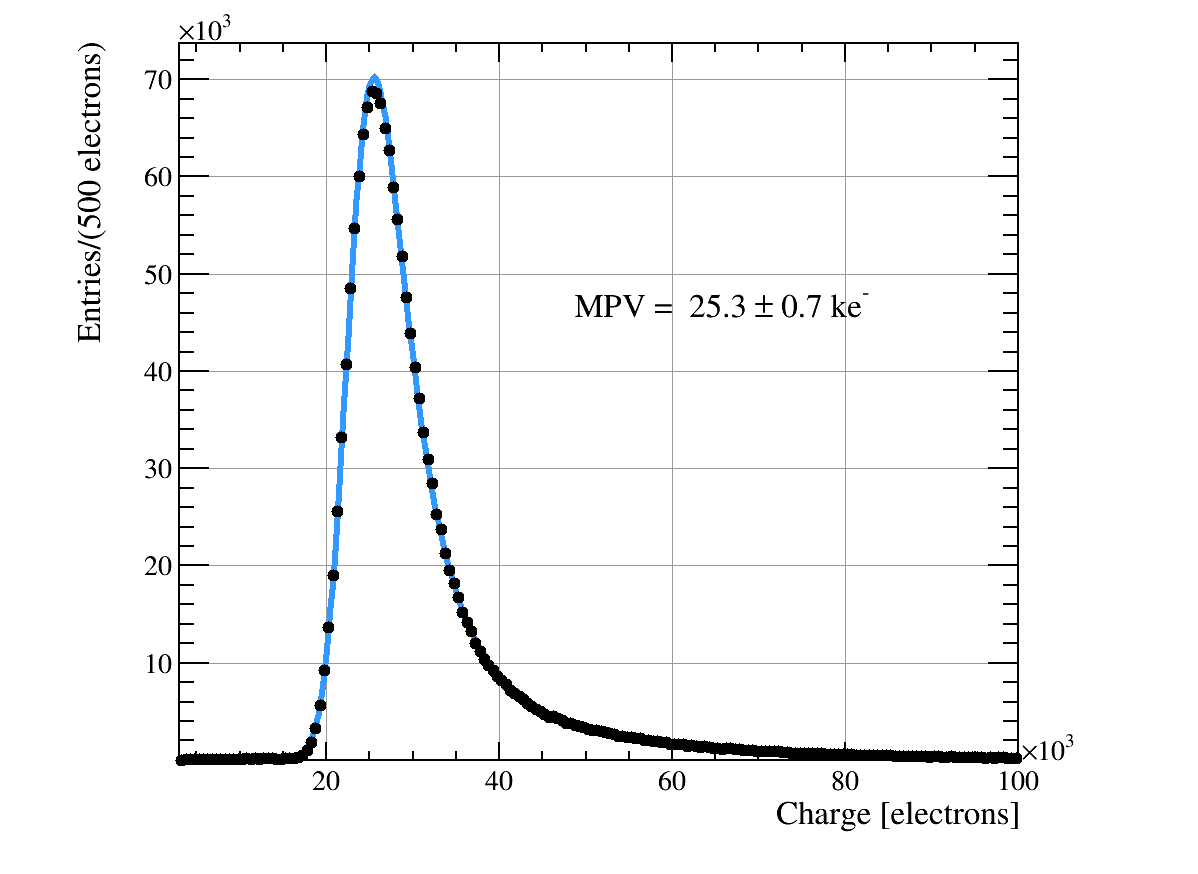}
  \caption{Typical distribution of collected charge of associated clusters for a telescope plane. 
The curve shows the Landau convoluted with a Gaussian fit.
} 
\label{Fig:PlanesCharge} 
\end{figure}

The charge collected in a typical telescope plane for clusters associated to a track is shown in
figure~\ref{Fig:PlanesCharge}.  
After applying the charge calibration described above, 
the distribution of charge deposited is well described
by a Landau function convoluted with a Gaussian distribution to account for the 
intrinsic resolution of the charge measurement.
The
extracted Landau MPV (Most Probable Value)
is found to be $25.3\pm 0.7$~ke$^-$, which is compatible with the expected value of a MIP crossing a 300~$\mum$
silicon layer.

\subsection{Time alignment}
\label{timealignment}

There are small time offsets between the planes of the telescope due to the time of flight of the particles
and non equal delays in the cables and electronics. 
Each plane is time aligned by minimising the variance of the cluster times (as defined in
section~\ref{cluster})  with a set of offsets $\tau_i$. 
This is equivalent to subtracting the mean of the biased track time residuals. 
The offsets are measured for all runs, and found to be consistent with each other to within a spread of about $15\ps$,
showing stability over time. 

\subsection{Clustering and tracking}\label{cluster}

Neighbouring pixel hits within a time window of 100~ns are
grouped into a cluster.
The cluster coordinates are calculated as the 
charge-weighted centre-of-gravity of the pixel hits constituting the cluster.
The difference in the charge-weighted and  ToT-weighted position
resolution was studied and found to be negligible.
The timestamp of the earliest hit in the cluster is used as the cluster timestamp, which mitigates the time
spread due to the timewalk of hits with low charge.

The pattern recognition is based on a simple track-following technique.
Track seeds are created from pairs of clusters on adjacent planes 
that are not yet part of a track, and differ by less than 10\,ns 
in their timestamps.
The seed tracks are extrapolated to the next plane following the beam propagation 
direction (downstream)  and the closest cluster in time  is added to the track 
if it is found within a temporal and spatial tolerance window.
This is repeated until all planes are searched.
If the candidate track comprises  8 clusters, 
it is accepted and a straight-line fit is performed.
The track time is given by the  arithmetic average of the timestamps of the clusters 
constituting the track. A list of the standard selection requirements used in the track reconstruction is shown in
Table~\ref{tab:stdcuts}.

\begin{table}
\begin{center}
\caption{List of the standard track reconstruction conditions.}
\begin{tabular}{ l |l| l}
\hline
\hline
Requirement                & Default value      & Description                                 \\
\hline
\hline
Cluster time window        &  < 100~\unit{ns}   & Maximum time difference of hits             \\
                           &                    & within the same cluster                     \\
\hline
Cluster width, in $x$ and $y$ &  $\le$ 3 pixels & Rejects large clusters from                 \\
                           &                    & $\delta$-rays, nuclear interactions etc.    \\
\hline
Track time window          &  $<$ 10~\unit{ns}  & Clusters within 10~ns window are            \\
                           &                    & considered for the pattern recognition      \\
\hline
Number of clusters         &  = 1 per plane     & Rejects multi prong interaction vertices    \\
in time window             &                    &                                             \\
\hline
Number of clusters         &  = 8               & Maximizes  track precision                  \\
per track                  &                    &                                             \\
\hline
Opening Angle              &  $<$ 0.01 rad      & Angle that defines the reconstruction       \\
                           &                    & window from plane to  plane,                \\   
                           &                    & assuming straight tracks                    \\
\hline
Fit $\chi^{2}$/ndof        &  $<$ 10            & Cut on $\chi^2$ divided  by the number of   \\
                           &                    & degrees of freedom for track quality        \\
\hline
\hline
\end{tabular}
\label{tab:stdcuts}
\end{center}
\end{table}

\subsection{Spatial alignment}
\label{alignment}

The planes of the telescope are aligned 
using the Millipede algorithm~\cite{Blobel2006}.
The principal advantage of this method is the simultaneous fit of 
both the track states and the geometry. 
The algorithm is iterated several times 
with progressively more restrictive selections on the tracks used.
A sample consisting of a fixed number of tracks, typically 8000, is reconstructed and an alignment tool
executed. This alignment is then used to update the geometry constants, and the next sample of tracks can be
reconstructed with reduced spatial windows and more restrictive requirements on track fit qualities.
This is repeated until either all alignment tools have been executed, or the data set has been exhausted.
All runs are automatically aligned using the same procedure. Alignment of DUTs that have a similar geometry
as the telescope planes is also provided by the Kepler software framework, and can be integrated into the automated
alignment sequence.


\section{Telescope Performance}\label{performance}
In this section the telescope performance results are presented.
The most relevant parameters assessed are the track time resolution (section~\ref{timeres}), the
spatial resolution 
(section~\ref{hitres}), together with its extrapolation to the track pointing resolution (section~\ref{pointres}).
Finally, an assessment of the high rate capabilities of the telescope is shown in section~\ref{highrate}.

\subsection{Time resolution}\label{timeres}
After the time alignment procedure described in section~\ref{timealignment}, the time resolution of
the telescope can be determined. The track timestamp is calculated as the average of the associated
cluster timestamps. 
The unbiased time resolution for a single plane is given by the standard deviation of  the time residual of the
clusters in one plane relative to the corresponding track, where the track time is the average time
of the other seven clusters on the track. 
This resolution is given by the combination of the  intrinsic time resolution 
of a single plane and the track resolution:
$$ \sigma^{2}_{\rm unbiased}  = \sigma^{2}_{\rm track} + \sigma^{2}_{\rm intrinsic}.$$
If the intrinsic time resolution ($\sigma_{\rm intrinsic}$)
is the same for each plane and the errors
on the time measurements are uncorrelated,
the track time resolution for a telescope with $N$ planes
is given by $\sigma_{\rm telescope} = \sigma_{\rm intrinsic}/\sqrt{N}$.
Since the unbiased resolution is measured on  a seven plane telescope,
$\sigma_{\rm intrinsic} = \sqrt{\frac{7}{8}} \sigma_{\rm unbiased}$. 

\begin{figure}[!htb]
    \centering
    \includegraphics[width=0.6\textwidth]{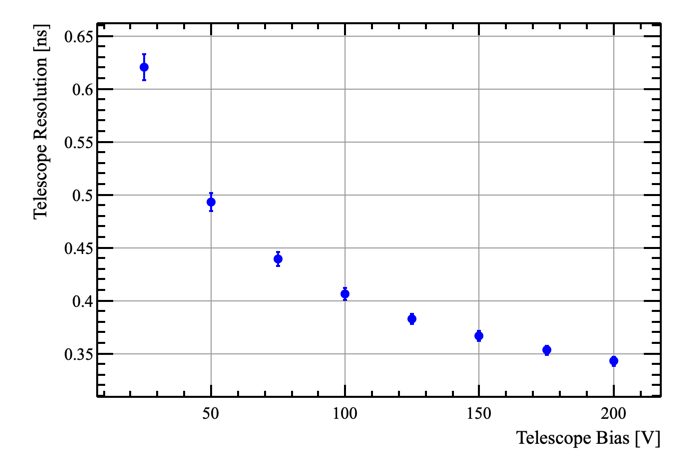} 
    \caption{Resolution of track timestamps as function of bias voltage,  using eight telescope
planes.}
    \label{fig:TimeResiduals}
  \end{figure}

The unbiased distribution in a 2$\times$2~mm$^2$ fiducial region shows a Gaussian shape with a width of $1.04\pm0.01$~ns and from this the intrinsic time
resolution is determined to be $\sigma_{\rm intrinsic} = 0.99\pm0.01$~ns per plane for a bias
voltage of 200 V. 
The difference between this result and the na\"\i vely expected value of $1.56 \ns / \sqrt{12}$ is
due to a mixture effects such as  timewalk in the chip, pixel-to-pixel variations,  the spread in charge collection times in the sensor, and
possibly  partial correlations of the time measurements in the different planes.
At 200 V bias the telescope time resolution from the combination of the eight uncorrelated measurements yields a
value of 0.35~ns. 
Figure~\ref{fig:TimeResiduals} shows the telescope time resolution as function of the 
applied bias tension to all planes, where this improvement is due to the lower charge collection
time. Increasing the bias voltage beyond 200 V might still yield a better        resolution but it
was considered to pose a risk of damaging the sensors.
This resolution was cross checked by measuring the time difference with respect to two scintillators
instrumented with constant fraction discriminators. 
The scintillator signals are timestamped with the TDCs of the SPIDR which have an average bin size of 260~ps and hence a resolution of about 75~ps. 
The time resolution of  the scintillator system is determined to be better than 0.3~ns and this
measurement confirms the track time resolution of $0.35$~ns.

\subsection{Spatial residuals}\label{hitres}

\begin{figure}[!htb]
\centering
\includegraphics[width=\textwidth]{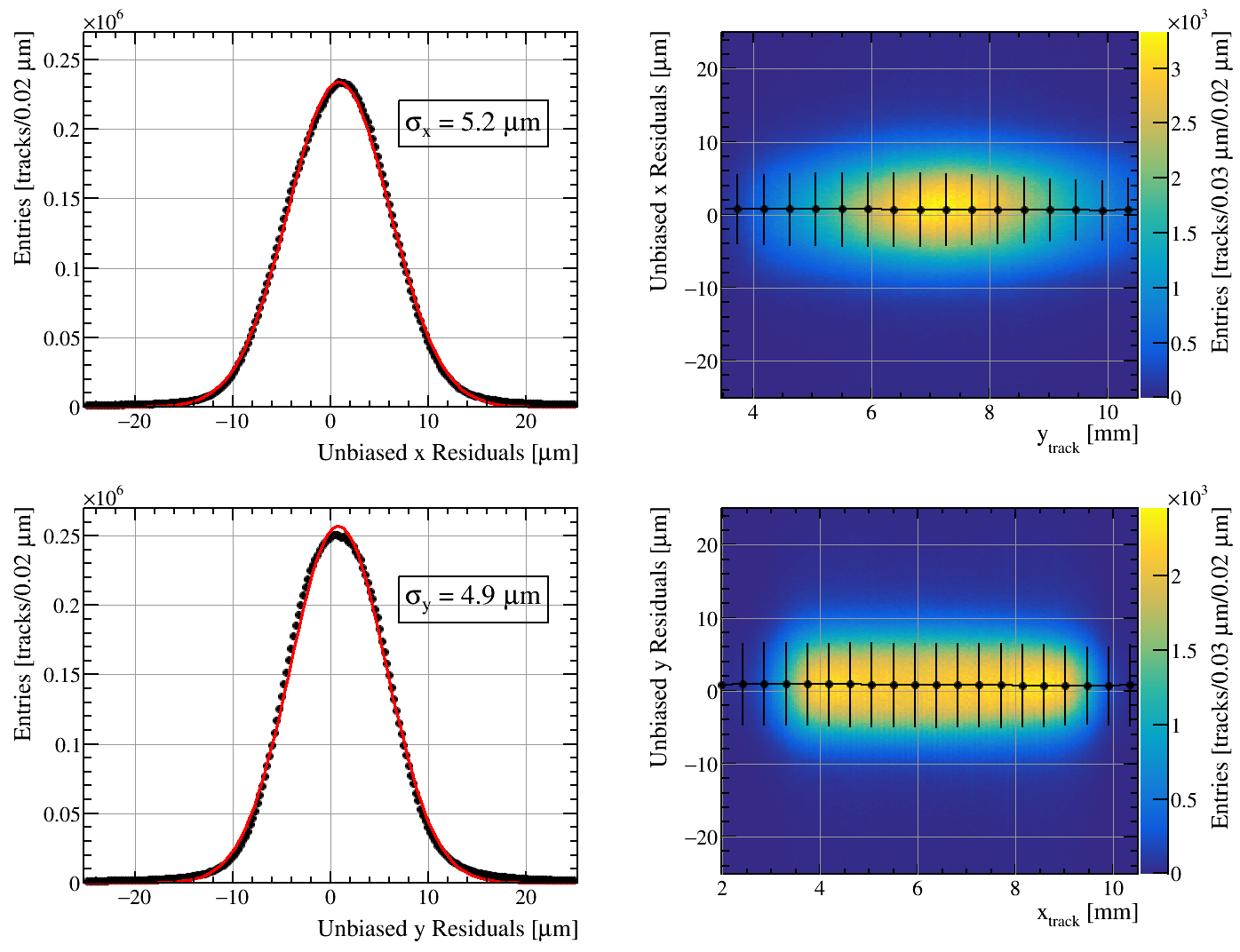} 
\caption{Unbiased residuals in $x$ and $y$ on one telescope plane, both integrated over the chip and as a
function of position. The error bars show the fitted gaussian width at each position.}
\label{fig:AlignmentResiduals}
\end{figure}

The precision of the spatial alignment of a plane is measured in an unbiased manner by excluding
it from the pattern recognition and
comparing the extrapolated position of tracks with the corresponding cluster on that plane.
These unbiased residuals are displayed for a typical plane in
figure~\ref{fig:AlignmentResiduals}, integrated over the sensor and as a function of position.   In
this case the residual is shown as a function of the orthogonal position, testing for rotations
around the z axis. 
Typically the variation across the beam spot relative to all axes is less than 0.2~\mum, and the
residual width is very stable.  The unbiased residual resolution is a convolution of the pointing
resolution of the telescope with the intrinsic resolution of the device.

The systematic error on the resolution of each plane is estimated from  the variation across the
plane in both $x$ and $y$.  A spread of 3-4\% (0.15~\mum) is obtained analysing the variations of
the standard deviation of the residuals distribution.
%

\subsection{Pointing resolution}\label{pointres}

A quantity which is of great interest for any device being tested in the telescope is the pointing
resolution; the precision with which the position of each track is known when  extrapolated to the
DUT position. For the Timepix3 telescope this resolution is a function of the measurement precision
of each plane, the spacing between the telescope planes, and the amount of material traversed by the
track. In the case of high energy ($>$ 10 GeV) hadronic beams the dominant contribution in this case is the 
measurement error of each plane.

In order to determine the track-pointing resolution a simulation procedure is used.  The first step
is to obtain an initial estimate of the cluster resolution of each individual plane.  Note that this
varies plane by plane due to the slightly different thresholds used and the variations in the angle
presented to the beam in both the $x$ and $y$ direction. The initial estimates are found by taking
the unbiased residuals and subtracting in quadrature a reasonable first guess of the track-pointing 
resolution at the $z$ position of that plane.  
The cluster resolution values so obtained are used as inputs to a fast
Monte Carlo simulation similar to the one described in Ref.~\cite{tpxtelepaper2}.   The simulation
gives as output the expected biased residual distributions at each plane, which are the
track-cluster residuals measured at each plane for tracks which use clusters from all 8 planes.  The
set of biased residuals obtained from the simulation are then compared to the data, and the input
errors are scaled by a single number for all planes until the best overall agreement with data is
achieved.   The cluster resolutions of the planes are found to be between 3.7 and 4.2~\mum in $x$
and 2.8 and 3.4~\mum in $y$.   The output of the simulation is used to derive the $z$-dependent
track-pointing error.  A conservative estimate of the error on this quantity is derived by varying
the  intrinsic spatial resolution of each plane by $\pm0.5$~\mum.

\begin{figure}[!htb]
\centering
\includegraphics[width=0.49\textwidth]{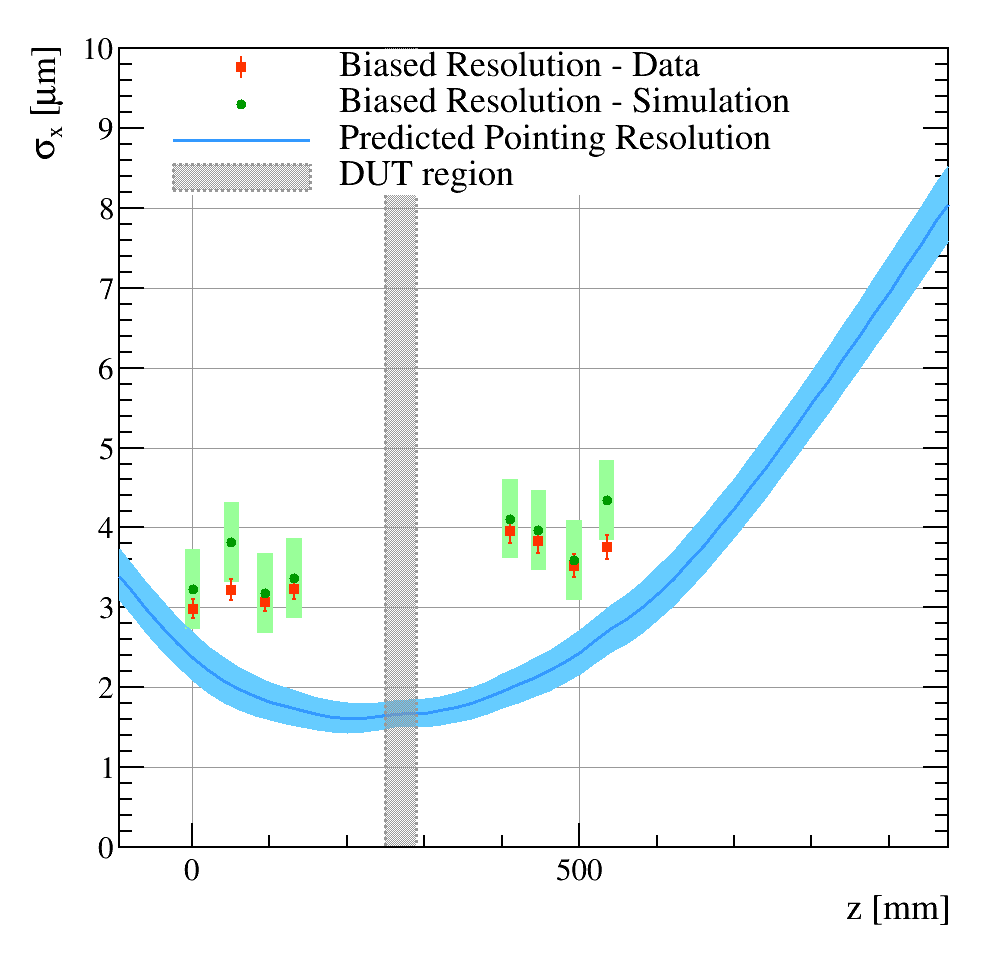}
\includegraphics[width=0.49\textwidth]{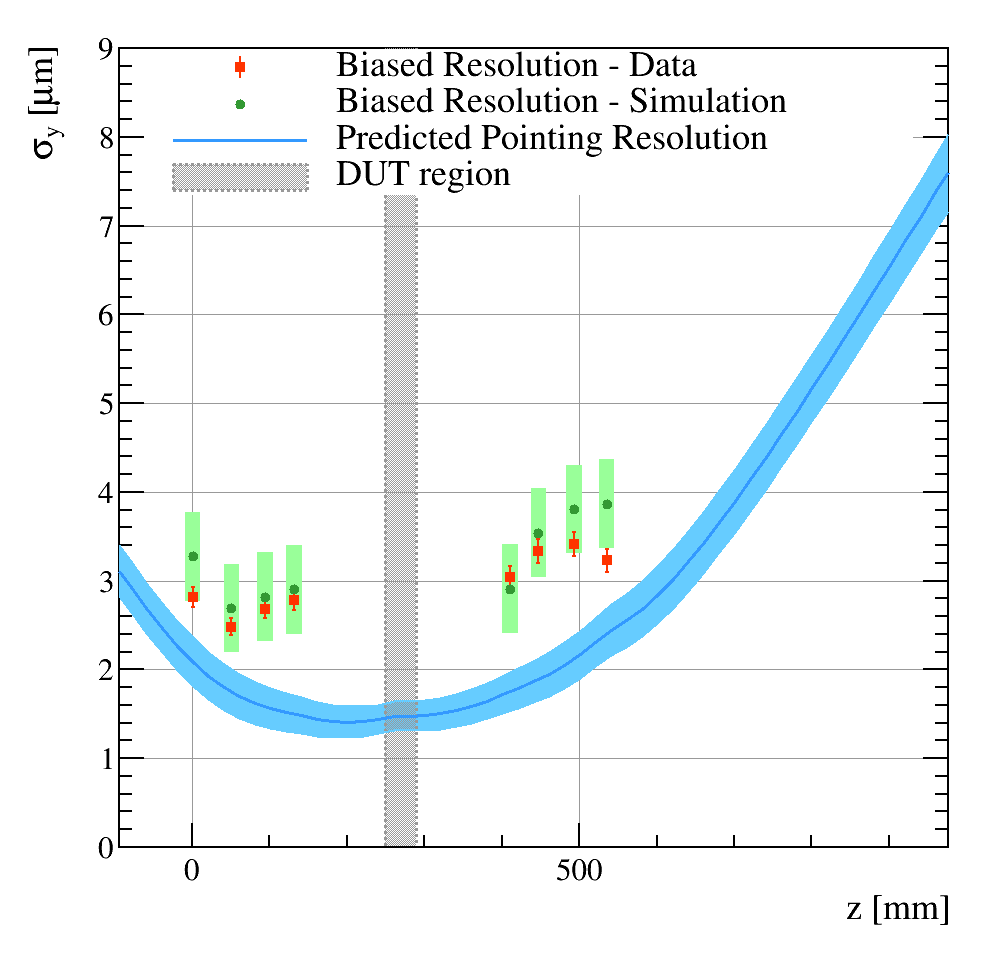}
\caption{
Predicted pointing resolution and uncertainty (solid blue curves) in the $x$~coordinate (left plot)
and $y$ coordinate (right plot).    The green dots show the Monte Carlo biased resolution and the
applied uncertainty is indicated by green bars. The red points show the biased resolution measured
in the data.  The typical installation position for a DUT is illustrated by the hatched grey 
region.  }
\label{fig:PointResolution}
\end{figure}

The result of the study is illustrated in figure~\ref{fig:PointResolution} for the $x$ and $y$
pointing resolution separately.  There is a good agreement between the biased residuals measured in
data and the ones predicted by the simulation. The predicted pointing resolution and uncertainty is
shown by the blue band.  The best pointing resolution is achieved at the DUT position in the centre
of the telescope.  Here the resolutions are $\sigma_{x} = 1.69\pm0.16$~\mum and 
$\sigma_{y}~=~1.55~\pm~0.16$~\mum.
 Also at the DUT stage downstream ($z\approx800$ mm) of the telescope a reasonable pointing precision
(< 10~\mum) is achieved.

\subsection{Track purity}

The track purity is defined as the fraction of reconstructed tracks that are 
due to a true charged particle traversing the telescope with respect to the total 
number of reconstructed tracks.
The precise time-stamping of each hit in the telescope allows a
very clean  pattern recognition.  Even at the highest rates of about ~$6.25\times 10^6$
particles/second, the average time between tracks is 16 times larger than the tracking time window,
and 160 times larger than the timestamp precision obtained in each of the planes.  The use of timing
measurements and the requirement to have exactly one cluster in each of the telescope
planes reduce the rate of fake tracks to a negligible level.  
The remaining contamination from fake tracks is estimated
by performing a fit to the spatial residuals of
the telescope including a flat background component. This component is taken as the upper bound
of the number of fake tracks and found to be below 2\textperthousand.  Hence the purity of the
tracks reconstructed by the telescope is above 99.8\%.

\subsection{High rate performance}\label{highrate}
\begin{figure}
  \centering
  \includegraphics[width=0.5\textwidth]{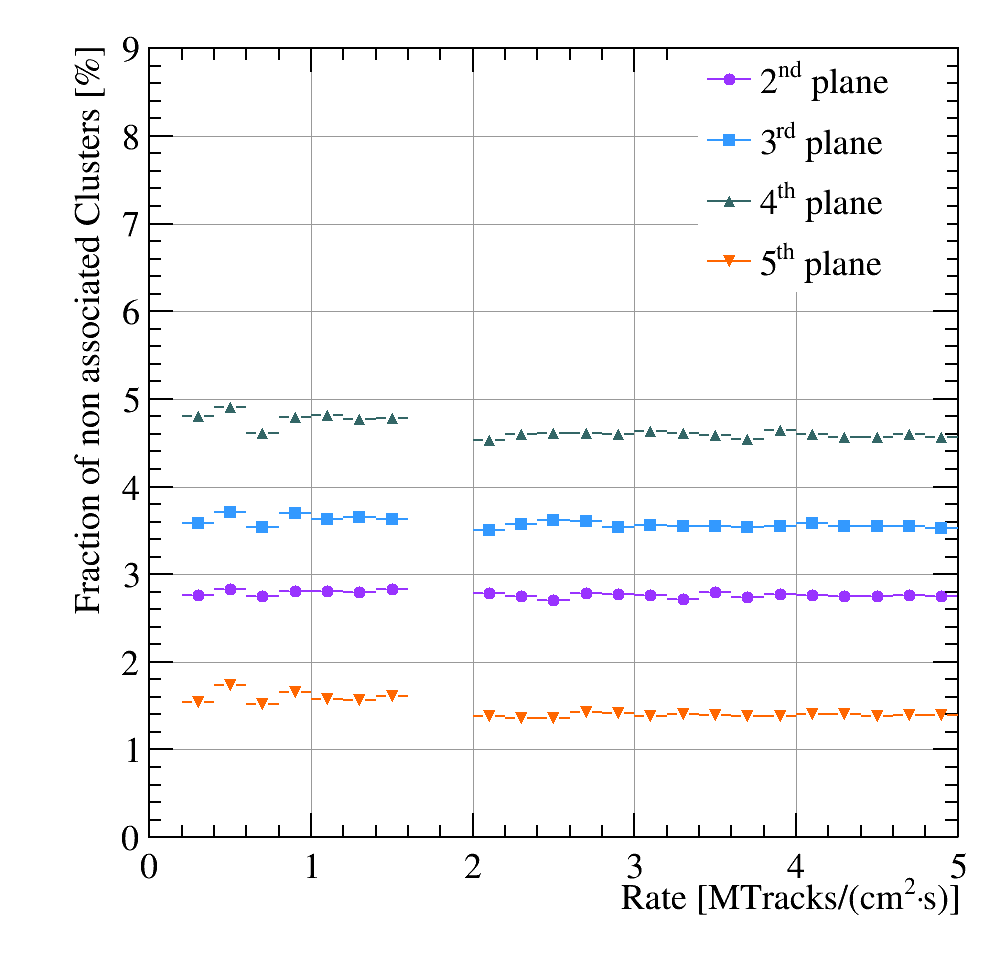}
  \caption{Fraction of clusters not associated to a reconstructed track in the internal planes of the
telescope. In the high-rate configuration the telescope was constituted by 6 planes only. The 
plane-to-plane variation does not depend on the rate and remains constant. This is attributed 
to slight differences among the ASICs.}
  \label{Fig:HR_performance}
\end{figure}


The beam intensity at the SPS experimental area can be modified by moving collimators located
upstream in the beam line.  The maximum achievable sustained intensity is $\approx 26$ million tracks
per spill of length 4.5~s.  This limitation is imposed by the background radiation alarms that dump
the beam once they are triggered.  In order to operate at high rates, the configuration of the
Timepix3 telescope was altered with respect to the one described in section~\ref{daq}.  The high-rate
configuration consists of six planes each read out by a dedicated SPIDR board to ensure enough
throughput to run the Timepix3 ASIC at its maximum rate.  During the dedicated high-rate performance
runs, the beam average particle rate was gradually increased from 70k to 26M tracks per 4.5~s spill.
Since the beam and the telescope are not exactly parallel, some particles will not traverse all
planes and hence not result in a reconstructed track.  To mitigate the impact of this effect, the
high rate performance is assessed in a fiducial region which contains the highest concentration of
reconstructable tracks.  

The fraction of clusters in each plane which are not associated to a track is studied in order to
verify that the telescope maintains a good pattern recognition efficiency at high track rates. The
dominant source of nonassociated clusters is particle interactions in the telescope material which
can generate $\delta$-rays, large angle scattering or vertices with multiple secondary tracks.  The
background rate from noise hits in the telescope is checked with runs  without beam and  found
to be negigible.  The fraction of nonassociated clusters should remain constant as the particle rate
increases, as long as the pattern recognition remains efficient and there is no significant increase
in the number of clusters on a particular plane.    This quantity is plotted in
figure~\ref{Fig:HR_performance} as a function of particle rate, where the rate is determined from the
telescope track reconstruction and timestamps.  Up to a rate of 5~MHz/cm$^2$, corresponding to the
maximum that the SPS can deliver, the rate of nonassociated clusters is constant, indicating that
the telescope track reconstruction efficiency is not significantly  degraded. 

\section{Summary}
A high speed telescope based on the \tpxt\ ASIC has been constructed. 
This paper presented the mechanics, data acquisition, slow controls, and performance figures.  
The simultaneous ToT and ToA measurements of the \tpxt\ offers a
fast, simple and robust pattern recognition and track reconstruction. 
The use of a charge-weighted clustering algorithm and a track-based alignment procedure provide residuals of the
order of 4~\mum for each telescope plane. 
The pointing resolution at the DUT position, in the centre of the telescope, is determined to be $1.69\pm0.16$~\mum.
A time resolution of 350 ps is achieved for reconstructed tracks traversing eight telescope planes. 
No deterioration of the telescope performance has been observed to a rate of  5~MHz/cm$^{2}$. 

\section{Acknowledgements}
We would like to thank the Medipix3 collaboration and especially Michael Campbell and Jerome Alozy
for providing the \tpxt assemblies for the telescope.  We also express our gratitude to our
colleagues in the CERN accelerator departments for the excellent performance of the PS and  SPS.  We
gratefully acknowledge the financial support from CERN and from the national agencies: CAPES, CNPq,
FAPERJ  (Brazil); the Netherlands Organisation for Scientific Research (NWO); The Royal Society and
the Science and Technology Facilities Council (UK).

\addcontentsline{toc}{section}{References}
\bibliographystyle{LHCb}
\bibliography{main}

\end{document}